# Detection and manipulation of surface electric field noise of hexagonal boron nitride

Hao-Jie Zhou[1,†], Xiao-Wen Shen[2,3,†], Yu Zhou[4,†], Lei Dong[2,3,†], Pei-Qin Chen[5,6], Xia Chen[1], Guang-Wei Deng[5,6], Gang Xiang[1], Pei-Jie Guo[1], Tian-Ke Wang[1], Hong-Peng Wu[2,3*], Jun-Feng Wang[1,*]

[1]College of Physics, Sichuan University, Chengdu 610065, China

[2]State Key Laboratory of Quantum Optics Technologies and Devices, Institute of Laser Spectroscopy, Shanxi University, Taiyuan 030006, China

[3]Collaborative Innovation Center of Extreme Optics, Shanxi University, Taiyuan 030006, China

[4]Ministry of Industry and Information Technology Key Laboratory of Micro-Nano Optoelectronic Information System, Guangdong Provincial Key Laboratory of Semiconductor Optoelectronic Materials and Intelligent Photonic Systems, Harbin Institute of Technology, Shenzhen 518055, China

[5]Institute of Fundamental and Frontier Sciences, University of Electronic Science and Technology of China, Chengdu 610054, China

[6]Key Laboratory of Quantum Physics and Photonic Quantum Information, Ministry of Education, University of Electronic Science and Technology of China, Chengdu 611731, China

† These authors contributed equally: Hao-Jie Zhou, Xiao-Wen Shen, Yu Zhou, Lei Dongs

* Corresponding author: wuhp@sxu.edu.cn, jfwang@scu.edu.cn

**Abstract**

**Hexagonal boron nitride (hBN) spin defects offer transformative potential for quantum sensing through atomic-scale proximity to target samples, yet their performance is fundamentally limited by rapid coherence loss. While magnetic noise mechanisms have been extensively studied, another critical influence from surface electric field noise remains unexplored in hBN systems. Here, we address**

**this challenge and systematically investigate surface electric field noise in hBN using shallow boron vacancy defects. The double-quantum spin relaxation behavior in response to magnetic fields and defect depths is examined, revealing that the relaxation rate follows a distinctive depth-related power-law dependence of ODMR splitting frequency. The relaxation is also demonstrated to be independent of the defect concentrations. Furthermore, the temperature dependence of the relaxation rate is investigated, showing a noticeable rise as the temperature increases from 296 K to 453 K, thus highlighting the influence of thermal effects on spin relaxation. To further suppress surface electric field noise, we explore the effectiveness of passivation materials, including glycerol and PMMA. Notably, PMMA is more efficient in mitigating surface electric field noise. These experiments enhance the understanding of surface electric field noise in hBN and provide a foundation for developing noise mitigation strategies in future research.**

**Introduction**

Optically active spin defects in solid-state systems such as diamond[1], silicon carbide[2] and hexagonal boron nitride (hBN)[3,4] have become leading platforms in quantum photonics, quantum communication, quantum computing, and quantum sensing[1-5]. Unlike three-dimensional diamond and silicon carbide, two-dimensional (2D) hBN is easily processable and integrable multifunctional heterostructures[6,7]. Since the optically detected magnetic resonance (ODMR) of the negatively charged boron vacancy ($V_B^-$) defects was discovered[4], the spin defects in hBN have attracted much attention in quantum technology[6-9]. The $V_B^-$ defect consists of a boron atom vacancy and an additional electron in the hBN crystal, and its direction is along the c-axis of hBN. The photoluminescence (PL) spectrum is about 700-900 nm. Similar to nitrogen-vacancy (NV) center in diamond, the spin state can be initialized and polarized by a 532 nm laser and controlled by microwave at room temperature[4,6,7]. The ODMR contrast can reach about 46% using a gold film microwave structure[10]. Due to its excellent optical and spin properties, it has been applied in quantum photonics[11,12], nuclear spin polarization and control[13,14], and quantum sensing including magnetic field[12], temperature[15,16], electric field[17], strain[18] and so on.

Besides these, one of the important quantum sensing is using shallow surface defects to sense external magnetic materials. Since the magnetic signal is inversely proportional to the cube of the distance between the spin defects and the target samples, decreasing the distance can effectively improve sensitivity and spatial resolution. Benefiting from the layered properties of two-dimensional hBN, the defects can be positioned at a very shallow distance from the surface[6,7,19,20,21]. In contrast to diamond surface, hBN surface does not contain the dangling bonds, which is benefit to form high-quality spin sensor. In view of these advantages, the shallow $V_B^-$ defects have been applied in sensing external paramagnetic spins[19,20] and stray fields of 2D magnetic materials[22]. Despite significant advancements, their relatively short spin coherence times ($T_2$ is about 50 ns to 2 μs and $T_1$ is about 5 - 20 μs at room temperature) limit its sensitivity in quantum sensing[6-10, 19-22]. Previous studies on shallow NV centers have demonstrated that the coherence time is limited by both magnetic field and electric field noise from the surface[23-25]. However, while significant attention has been given to magnetic noise from the electric and nuclear spin bath in $V_B^-$ defects in hBN, the impact of surface electric field noise remains largely unexplored. Therefore, investigating electric field noise is crucial for gaining a comprehensive understanding of the decoherence mechanisms in shallow $V_B^-$ defects in hBN, which is also fundamental to the advancement of quantum sensing.

In this work, we systematically study the surface electric field noise of hBN using shallow $V_B^-$ defects by double-quantum (DQ) relaxation method. By investigating the DQ relaxation rate $\gamma$ under varying magnetic fields, defect depths, and defect concentrations, we reveal that $\gamma$ exhibits a depth-related power-law dependence of ODMR splitting frequency, which is independent of the defect concentrations. Furthermore, we extend our measurements over a broad temperature range from 296 K to 453 K to explore the thermal behavior of $\gamma$. Finally, we manipulate the surface electric field noise using a passivation layer (PMMA and glycerol), providing insights into noise mitigation strategies for improving spin coherence properties in hBN. These experiments lay the groundwork for understanding surface electric field noise, which is crucial for its applications in quantum sensing.

**Results and discussion**

In the experiments, several single crystal hBN samples (HQ Graphene) are used, and shallow $V_B^-$ defects are created by implanting nitrogen ions at 2.5 keV with a dose of $2\times10^{14}$/cm$^2$. SRIM simulations indicate that the average defect depth is approximately 4.8 nm. The ground-state spin Hamiltonian of the $V_B^-$ center can be expressed as:

$$H = D\left(S_z^2 - \frac{S(S+1)}{3}\right) + E(S_x^2 - S_y^2) + g\mu_B \mathbf{B}\cdot\mathbf{S} \tag{1}$$

where **S** is the electronic spin operator (S=1), $D_{gs} \approx 3.48$ GHz and $E_{gs} \approx 48$-75 MHz are the ground state zero-field splitting (ZFS) parameters, $g$ is the Landé $g$-factor ($g = 2$), and $\mu_B$ is the Bohr magneton. For the excited state, the ZFS parameters $D_{es} \approx 2.1$ GHz and $E_{es}$ is about 93-154 MHz[6,7]. The last term describes the Zeeman splitting, accounting for energy level splitting in the presence of an external magnetic field, with the splitting energy given by $\hbar\omega_\pm = 2g\mu_B B$. Figure 1A shows the schematic of detecting surface electric field noise using shallow $V_B^-$ defects, due to unpaired electrons, moving charges and fluctuating dipoles[25, 26].

The $V_B^-$ defects energy level diagram is illustrated in Fig. 1B. Considering the three-level system of the $V_B^-$ defects, DQ relaxation between |+1⟩ and |-1⟩ states is sensitive to electric field noise and single-quantum (SQ) relaxation between |0⟩ and |±1⟩ states is sensitive to magnetic field noise[23-25]. The DQ and SQ relaxation rates are denoted by γ and Ω, respectively. Assuming complete relaxation within the three-level system, the population dynamics of each state can be described by the rate equations, which are similar to NV centers in diamond[23-25]. The relaxation rates Ω and γ can be extracted using the experimental pulse sequence illustrated in Fig. 1C. Laser pulses are used to polarize and read out the spin state of the $V_B^-$ defects, while microwave $\pi_\pm$ pulses denote transitions between |0⟩ and |±1⟩ states. By fitting the resulting decay curves, the SQ and DQ relaxation rates can be determined using the following equations:

$$F_1 = re^{-3\Omega\tau} \tag{2}$$

$$F_2 = re^{-(2\gamma+\Omega)\tau} \tag{3}$$

where τ denotes the evolution time, similar to the results of NV centers in diamond[23-25]. Figure 1D shows the measurements of DQ and SQ relaxation of $V_B^-$ defects (depth 4.8 nm) under a c-axis external magnetic field (14 G), with DQ relaxation shown as red circles and SQ decay shown as blue squares. Solid lines represent the fitted curves using Equations (2) and (3), yielding γ = 99.8 ± 11.9 kHz and Ω = 35.1 ± 2.7 kHz.

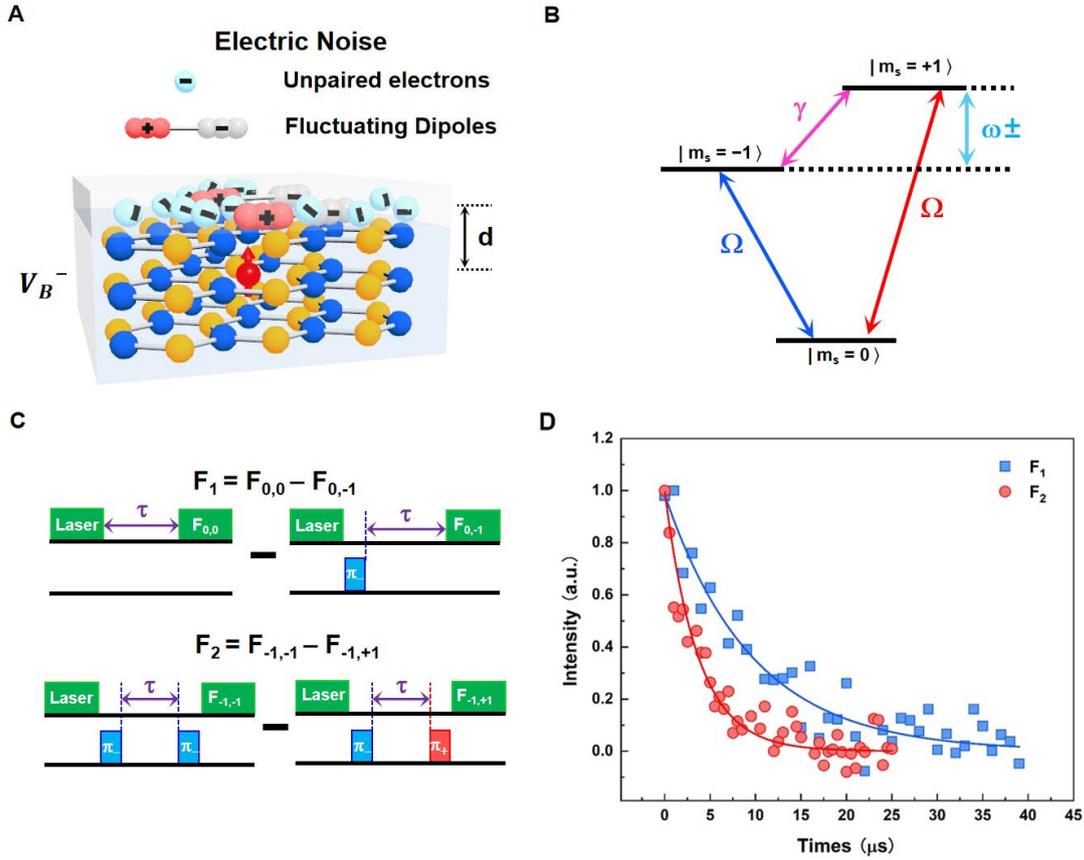

Fig. 1. Schematic and pulse sequences of detection of surface electric field noise using shallow $V_B^-$ defects. (A) Schematic of the $V_B^-$ defects in hBN and surface electric field noise from unpaired electrons, moving charges and fluctuating dipoles. (B) Energy level diagram of the ground-state triplet of $V_B^-$ defects. γ represents the DQ relaxation rate between |±1⟩ states, Ω represents the SQ relaxation rate between |0⟩ state and |-1⟩ or |+1⟩ state. $\omega_\pm$ represents the energy splitting between |-1⟩ and |+1⟩ state. (C) Pulse sequences to extract the relaxation rates Ω and γ. Green squares represent laser pulses for polarize and reading the spin state of $V_B^-$ defects. Blue squares represent $\pi_-$ pulses between the |0⟩ and |-1⟩ state, while the red ones denote $\pi_+$ between |0⟩ and |+1⟩ state. (D) Experimental results of DQ (red circles) and SQ (blue squares) of $V_B^-$ defects (depth 4.8 nm) under a c-axis magnetic field. Solid lines are fits to

Equation (2) and (3).

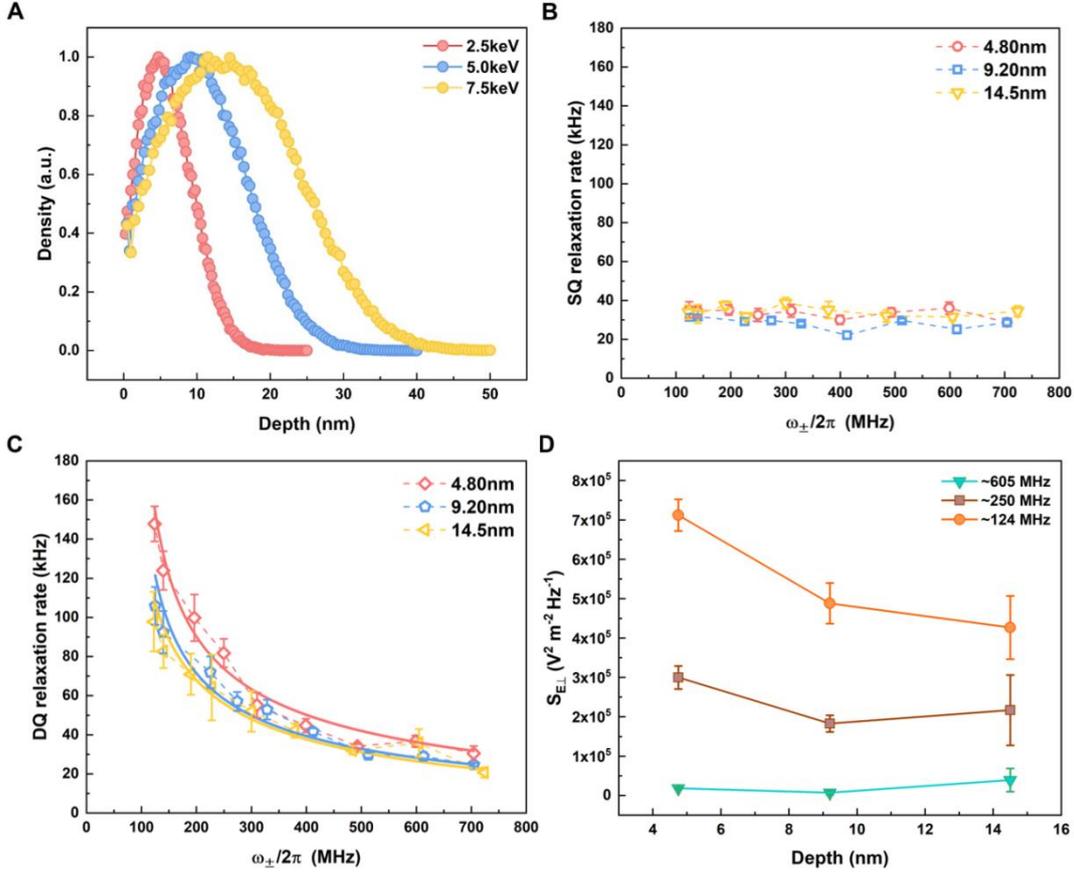

**Fig. 2. Depths dependence property of the DQ relaxation.** **(A)** Depth profile of $V_B^-$ defects created by implanting 2.5, 5 and 7.5 keV nitrogen ions, with most probable depths of 4.8, 9.2, and 14.5 nm, based on SRIM simulations. **(B)** SQ relaxation rates at varying depths under different external magnetic fields. **(C)** DQ relaxation rates of different defect depths as a function of magnetic fields. Solid lines represent data fits. **(D)** Comparison of electric field noise intensity $S_{E_\perp}$ of different defect depths at different energy splitting. Error bars represent one standard deviation from the fit.

Considering the electric field noise at the surface of hBN and the potential influence of the defect depth on the relaxation rate of $V_B^-$ defects, we initially explored the depth-dependent behavior of the SQ and DQ relaxation rate in $V_B^-$ defects. We generate $V_B^-$ defect ensembles at three different depths by varying the nitrogen ion implantation energy, while keeping the ion dose constant. According to SRIM simulations, the most probable defect depths are approximately 4.8 nm, 9.2 nm, and 14.5 nm, corresponding to implantation energies of 2.5 keV, 5 keV, and 7.5 keV (Fig. 2A). Figure 2B presents the measured SQ relaxation rates under varying external magnetic fields. The SQ rate remained unchanged across different defect depths, indicating that Ω is independent of the $V_B^-$ defect depth, similar to previous results[10]. Figure 2C shows the DQ relaxation rates $\gamma$ as a function of magnetic field strength for defects at different depths. Consistent with previous studies, $\gamma$ can be fitted by the function $1/(f - 2E)^a + \gamma_\infty$, where $f$ is ODMR splitting[23,24]. The component of $\gamma_\infty$ is attributed to the contribution of bulk effect, and the $1/(f - 2E)^a$ part is ascribed to the surface electric-field noise[23,24]. The solids lines are the fitting using the function. The experiments show that all three DQ rates $\gamma$ decrease with increasing magnetic field strength and then approach saturating for magnetic field of $\omega_\pm/2\pi$ > 500 MHz, which are similar to the results of NV centers in diamond[23-25].

The noise intensity of the electric field component perpendicular to the c-axis can be described by the following equation[23,24]:

$$S_{E_\perp} = \frac{\gamma - \gamma_\infty}{(d_\perp/h)^2} \tag{7}$$

Where $d_\perp/h$ = 0.4 Hz m/V represents the ground-state transverse electric field susceptibility of the $V_B^-$ defects[27]. Figure 2D compares electric field noise intensity of different defect depths at different energy splitting. At low magnetic fields, the shallow defects demonstrated significantly higher $S_{E_\perp}$ values compared to deeper defects, whereas this difference diminished at higher fields. These results indicate that surface-related electric field noise is the primary source of DQ relaxation in hBN, and the influence to DQ relaxation increases as the $V_B^-$ defect depth decreases.

Having established the critical role of defect depth in surface electric field noise, we next investigate whether defect-defect interactions contribute to the observed relaxation dynamics by varying the implantation dose while maintaining the defects depth. Nitrogen ions are implanted with an energy of 2.5 keV, while the implantation doses ranging from 2×10$^{13}$ /cm² to 4×10$^{15}$ /cm². As the implantation dose increased, the $V_B^-$ defects PL initially increases but showed a slight decrease when the dose exceeded 2 ×

$10^{14}$/cm² (Supplementary Materials). This trend is likely due to the saturation effect of $V_B^-$ defects, which is consistent with previous results[28].

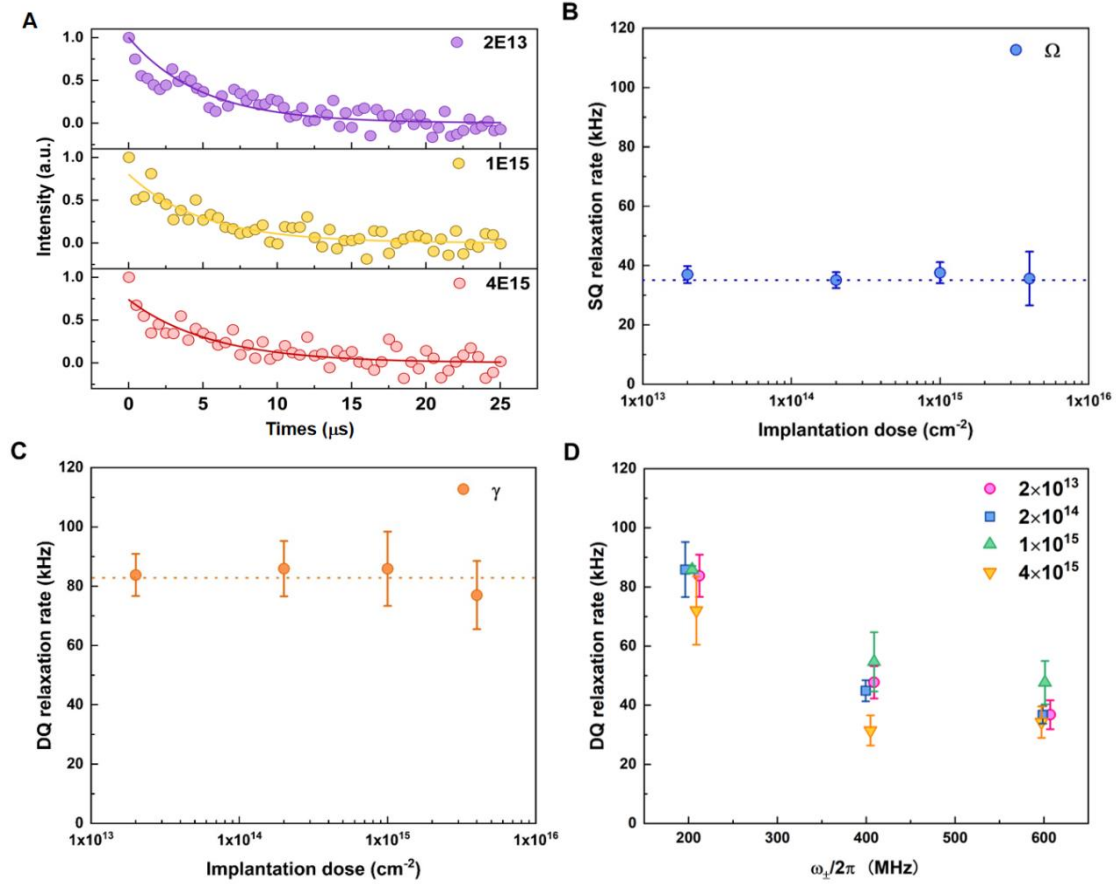

**Fig. 3. DQ relaxation dynamics with varying defect concentrations. (A)** Three measurements of $\gamma$ at different $V_B^-$ defect concentrations. The external magnetic field is ~ 18 G. Dots represent experimental data. Solid lines represent data fits. **(B)** Summary of SQ and **(C)** DQ relaxation rates at different concentrations. Dots are experimental data, and dashed lines are guides to the eye. Both $\gamma$ and $\Omega$ are independent of implantation fluences. **(D)** DQ relaxation rates at different defect concentrations for three energy splitting. Error bars represent one standard deviation from the fit.

Figure 3A presents representative measurements of $\gamma$ for samples with different implanted doses under a magnetic field of approximately $\omega_\pm / 2\pi \approx 200$ MHz. For implantation doses of $2 \times 10^{13}$, $1 \times 10^{15}$, and $4 \times 10^{15}$/cm², the measured $\gamma$ values were $83.7 \pm 7.1$ kHz, $85.8 \pm 12.5$ kHz and $77.0 \pm 11.5$ kHz, respectively. Figure 3B shows that the SQ rate $\Omega$ is independent of the $V_B^-$ concentration[29]. Figure 3C illustrates the

DQ rate $\gamma$ as a function of defect concentration, revealing that $\gamma$ is also independent of the concentration, consistent with results from NV centers[29]. Figure 3D shows $\gamma$ measured at different defect concentrations under three magnetic fields. $\gamma$ remains nearly constant across concentrations at the same magnetic field, indicating that defect concentration does not affect $\gamma$ and supporting the conclusion that the DQ relaxation rate $\gamma$ is governed by surface electric field noise[29].

We then extend our investigation to temperature-dependent dynamics across 296 – 453 K, probing the interplay between phonon interactions and surface noise. As the temperature increased, the ODMR peak decreases, which is attributed to the reduction of the ZFS parameter $\mathbf{D}$[16] (Supplementary Materials). Previous studies observed a monotonic increase in the longitudinal spin relaxation time $T_1$ at lower temperatures (20-300 K)[30, 31]. However, those studies primarily relied on the conventional definition of $T_1^{(0)}$, expressed as $T_1^{(0)} = (3\Omega)^{-1}$, which neglects the contribution of $\gamma$[23]. Given the nonzero contribution of $\gamma$ observed in our experiments, a revised definition of the spin relaxation time $T_1$ is required for a more accurate description:

$$\frac{1}{T_1} = \frac{1}{T_1^{(0)}} + \gamma = 3\Omega + \gamma \qquad (4)$$

which is similar to NV centers in diamond[23].

The temperature dependence of $\gamma$ is examined by measuring SQ and DQ relaxation rates under a magnetic field of 36 G. Figures 4A and 4B show the respective SQ and DQ relaxation results under three different temperatures, respectively. At room temperature, the values of $\Omega$ and $\gamma$ were measured to be 40.73 ± 1.36 kHz and 88.26 ± 8.1 kHz. As the temperature increased 453 K, these values rose to 112.54 ± 11.78 kHz and 255.6 ± 58.27 kHz. Figure 4C illustrates the temperature dependence of $\Omega$ and $\gamma$. As the temperature increased, $\gamma$ exhibited a significant rise, which is due to the increased phonon bath[24]. The increased uncertainty is due to the reduced contrast of the ODMR signal at higher temperatures. The spin-lattice phonon interactions become stronger with increasing temperature, leading to a more substantial contribution from $\gamma_\infty$ at elevated temperatures[24]. At the same time, the SQ relaxation rate $\Omega$ also increases significantly with temperature. Considering the complete state dynamics, the conventional definition of the relaxation time $T_1^{(0)}$ tends to overestimate the full spin relaxation time. Figure 4D provides a quantitative analysis of this temperature dependence, with the inset displaying a double logarithmic plot of the spin-lattice relaxation rate $1/T_1$. A linear fit yields a slope of 2.44, indicating that $1/T_1 = 3\Omega + \gamma \propto T^{2.44}$. This result is consistent with the slope of 2.5 previously observed at lower temperatures[30].

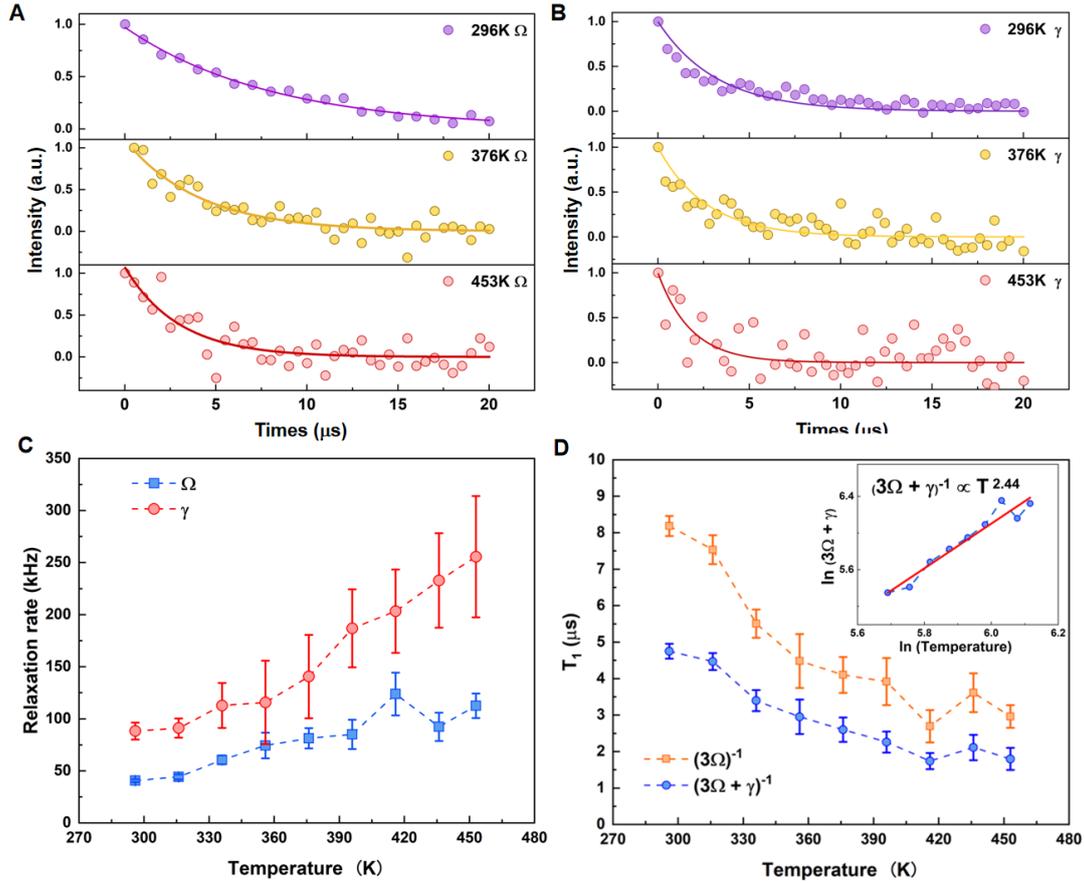

**Fig. 4. Temperature dependence of DQ relaxation rates over the range of 296 K to 453 K. (A) and (B)** SQ and DQ relaxation measurement at different temperatures. The external magnetic field is ~ 36 G. Dots represent experimental data, and the solid lines are fits. **(C)** DQ and SQ relaxation rate as a function of temperature from 296 K to 453 K. Error bars represent one standard deviation derived from the fit. **(D)** Comparison of the temperature-dependence of traditional spin relaxation time $T_1^{(0)}$ and full $T_1$. The inset shows a log-log plot for $1/T_1$, revealing $T^{2.44}$ behavior. Red line is a linear fit.

A key challenge in extending the coherence time of spin defects in hBN is suppressing surface electric field noise, which accelerates spin relaxation and limits quantum coherence[23-25]. Dielectric materials including PMMA and Glycerol have been chosen as capping layers in shallow NV centers due to their known ability to suppress relaxation[24,25,32].

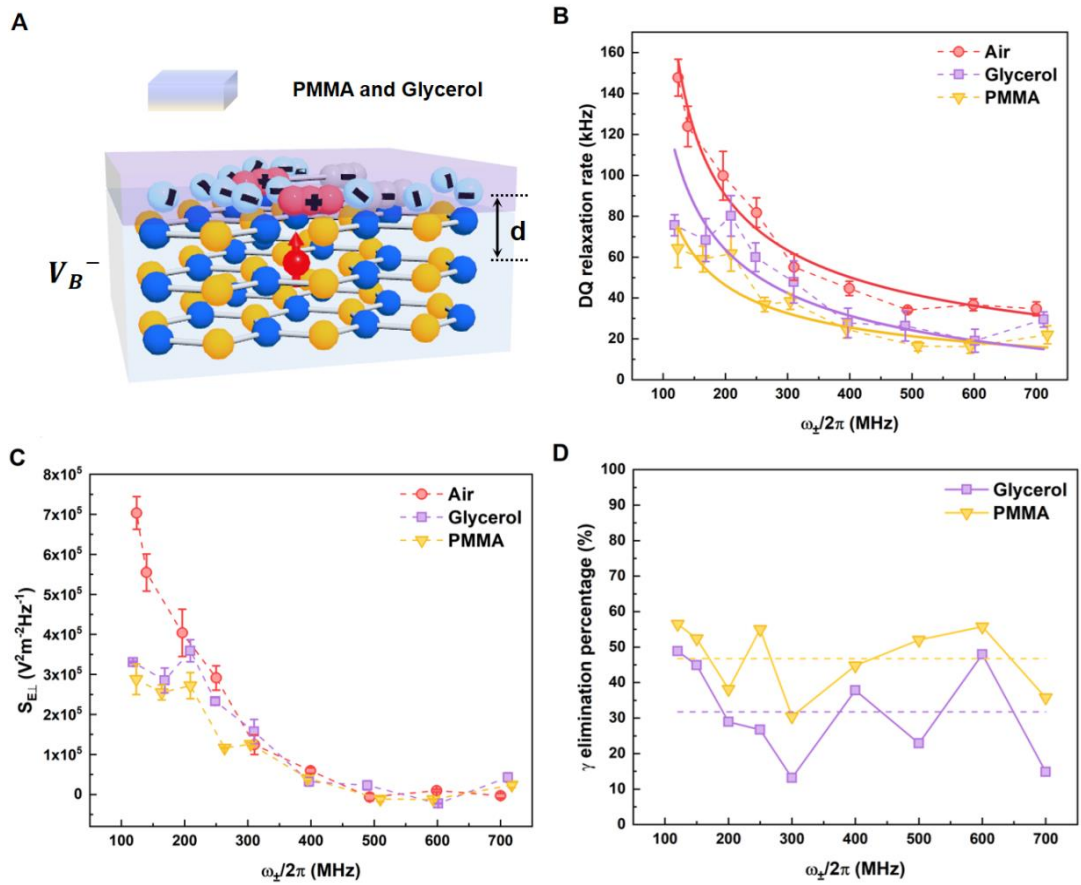

**Fig. 5. Manipulation of surface electric-field noise using PMMA and glycerol covering layers.** **(A)** Schematic of hBN covered by PMMA and glycerol covering layers, with the shallow $V_B^-$ defects (d = 4.8 nm). **(B)** Comparison of $\gamma$ with and without covering layers. Solid lines represent data fits, showing consistent power-law frequency dependence. **(C)** Electric field noise spectra with PMMA and glycerol covered on the surface. **(D)** Percentage of electric noise suppression achieved by PMMA and glycerol compared to raw hBN. Average suppressions over the frequency range are 46.7% for PMMA and 31.8% for glycerol. Dashed lines are guides to the eye.

To explore the effectiveness of these capping layers in noise suppression in hBN,

we examined their influence on shallow $V_B^-$ defects (depth 4.8 nm). Figure 5A presents a schematic of shallow $V_B^-$ defects with glycerol and PMMA applied as capping layers. Figure 5B shows the measured $\gamma$ for hBN samples under different capping conditions, and all the $\gamma$ exhibited a power-law frequency dependence, confirming that the relaxation process is due to the surface electric field noise[23,24]. Figure 5C presents the surface electric field noise spectral density, revealing that both glycerol and PMMA effectively can effectively reduce electric field noise, and PMMA has a better effect, which is similar to the previous experiments in NV centers. Due to its higher dielectric constant ($\kappa_{glycerol}$= 42), glycerol demonstrates substantial noise suppression across a broader frequency range[24,25,32]. Interestingly, unlike the behavior observed in shallow NV centers[24], the noise suppression effect at broad frequency splitting does not diminish. This variation may reflect a difference in the origin of surface electric field noise between hBN and diamond[6,7]. Figure 5D further compares the noise suppression efficiency of glycerol and PMMA. The PMMA achieves an average noise suppression of 46.7% compared to 31.8% for glycerol across the experimental frequency range. The superior suppression observed with solid state PMMA may be attributed to surface-modified phonon coupling effects[24,25,32,33].

**Conclusion**

In conclusion, the work systematically investigates and manipulates the surface electric field noise of hBN using shallow $V_B^-$ defects through the DQ relaxation method. Our results reveal that the $\gamma$ relaxation rate follows a power-law frequency dependence, indicating surface-dominated noise mechanisms. To further explore the origin of $\gamma$, we examined its correlation with the depth and defects concentration. The analysis demonstrated that the electric field noise decreases as the defect depth increases, and defect concentration doesn't affect the noise. We then investigate the temperature dependence of $\gamma$ in the range of 296 K to 453 K, observing a significant rise due to the increased phonon bath. Finally, we use the passivation layers, including solid PMMA and liquid glycerol, to effectively suppress surface electric field noise, and PMMA has a better effect (46.7%) across a broad frequency range. This study provides comprehensive insights into the properties and suppression methods of surface electric field noise in hBN, contributing to a deeper understanding the influence in the decoherence mechanisms of the $V_B^-$ defects. Our findings offer a solid foundation for the continued development and application of $V_B^-$ defects in quantum sensing of various magnetic materials[19-22,34-37] and are also applicable to other 2D spin defect systems[8,35,38].

## Methods

### Samples and experimental setup

To eliminate the influence of the substrate on the $V_B^-$ defects, all hBN samples used in this work are thicker than 100 nm. The experimental setup is consist of a custom-designed room-temperature confocal microscope system equipped with a 532 nm excitation laser and a single-photon detector to capture photoluminescence (PL) signals. A 650 nm long-pass filter was integrated into the system to filter out the excitation laser. The spin states of $V_B^-$ defects are manipulated through a microwave set up, and an external magnet is used to apply the c-axis magnetic field.

## Acknowledgements

We would like to thank Igor Aharonovich for helpful discussion. This work was supported by the National Natural Science Foundation of China (Grant No. 12474499), the Sichuan Science and Technology Program (Grant No. 2024ZYD0022).